\documentclass[letterpaper,10pt]{article} 

\usepackage{opticameet3} 
\usepackage{caption}
\usepackage{svg}

\newcommand\authormark[1]{\textsuperscript{#1}}

\usepackage{amsmath,amssymb}
\usepackage[colorlinks=true,bookmarks=false,citecolor=blue,urlcolor=blue]{hyperref} 

\begin{document}

\title{FPGA-Based Experimental Analysis of Fixed-Point Precision Impact on SOP Estimation in Coherent Communications Receivers}


\author{Geraldo Gomes\authormark{1},
Rafael Vieira\authormark{1},
Hani Kbashi\authormark{1},
Aleksandr Donodin\authormark{1},
Shekhar Saxena\authormark{1},
Stylianos Sygletos\authormark{1},
Ian Phillips\authormark{1},
Jaroslaw E. Prilepsky\authormark{1}, 
Mikael Mazur\authormark{2},
Sergei K. Turitsyn\authormark{1}, 
Pedro Freire\authormark{1,*}
} 

\address{
\authormark{1}Aston Institute of Photonic Technologies, Aston University,  B4 7ET Birmingham, UK\\
\authormark{2}Nokia Bell Labs, Murray Hill, NJ 07974, USA\\
}

\email{\authormark{*}freiredp@aston.ac.uk} 

\vspace{-7mm}

\hspace{-4mm}
\begin{abstract}
We experimentally evaluated the sensing-communication trade-off from the fixed-point precision MIMO equalizer using FPGA. At 7-bit, noise floor drops  $\sim$100× and angular error 63\%, but the communication performance saturates while the hardware complexity rises.
\end{abstract}
\vspace{-0mm}

\section{Introduction}
\vspace{-1mm}
Environmental sensing using state of polarization (SOP) variations derived from Jones matrices of live submarine optical links has recently gained attention \cite{Mecozzi2021, Zhan2025}. This technique enables simultaneous sensing and communication, eliminating the need for dedicated wavelengths or dark fibers \cite{Mazur2022}. Environmental perturbations induce fiber strain and birefringence changes, which can be characterized through Jones matrices estimated via a constant modulus algorithm (CMA) equalizer \cite{Kazuro2011} implemented for adaptive filtering as part of the digital coherent receiver's DSP chain. From these matrices, the SOP evolution can be reconstructed on the Poincaré sphere.

\begin{figure}[!b]
    \centering
    \includegraphics[width=.9\textwidth]{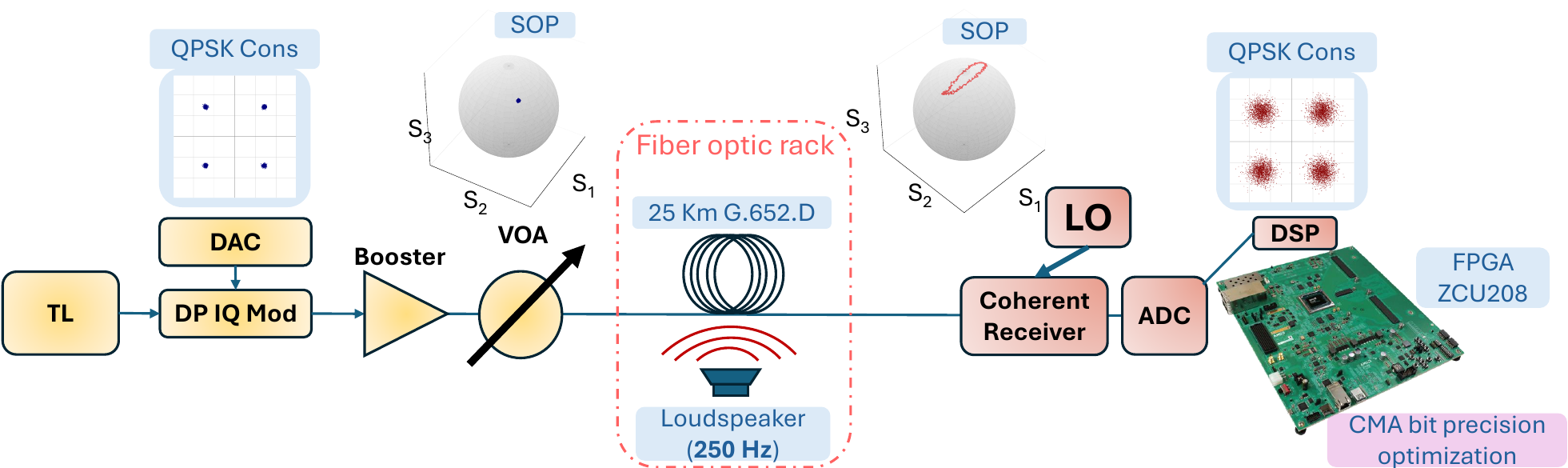}
    \captionsetup{font=footnotesize}
    \caption{Sketch of the experimental setup: DP-QPSK communication; input vibration from loudspeaker; DSP with CMA used for SOP vibration sensing; CMA bit precision optimization for sensing.}
    \label{Figura1_telling_story}
    \vspace{-2mm}
\end{figure}

While previous studies optimized CMA bit precision for communication performance \cite{Chin2012}, the sensing implications of the fixed-point precision used in the equalizer remain unexplored. Understanding the bit-precision impact is crucial, as it directly impacts area usage and power dissipation in ASIC implementations. In this work, we experimentally investigate how CMA bit precision influences SOP measurement accuracy in a 25 km optical link exposed to mechanical vibrations induced by a loudspeaker at 250Hz that induces polarization state rotations (Fig. \ref{Figura1_telling_story}). The received signal is processed offline, where only the CMA equalizer, responsible for polarization demultiplexing, is implemented in fixed-point precision on an FPGA to simulate realistic hardware constraints.
Our results show that sensing performance is highly sensitive to bit-width: increasing the CMA precision from 5 to 7 bits reduces the angular error by 63\% and lowers the noise floor by nearly two orders of magnitude ($\sim$100×). This demonstrates that SOP-based sensing demands higher bit-precision than conventional communication systems to ensure stable and low-noise polarization tracking. Additionally, we observe that increasing the bit-width also raises hardware complexity, leading to higher consumption of LUTs and flip-flops and a larger overall chip area in the FPGA implementation.

 \vspace{-2mm}

\section{Experimental Setup and Methods}
\vspace{-1mm}

The DP-QPSK experiment employed a 64 GHz oscilloscope to capture IQ data, a 25 km standard SMF spool for transmission, and a loudspeaker with built-in synthesizer positioned adjacent to the fiber to enhance mechanical coupling. Communication was performed with a dual polarization 10 GBaud QPSK signal with a root-raised-cosine (RRC) filter, roll-off of 0.1.
The 0dBm signal was launched into 25 km of standard single-mode fiber (ITU-T G.652.D) with total optical loss of 5 dB. The fiber was subjected to a 250 Hz sinusoidal vibration generated by the loudspeaker. 
At the receiver, coherent detection was performed using a standard intradyne coherent receiver with a free-running local oscillator (linewidth 100 kHz). The analog outputs were sampled by a 59 GHz oscilloscope with 64 GSa/s for offline processing.
The raw IQ data were processed offline with a DSP chain implemented in MATLAB, except for the MIMO CMA equalizer, which was implemented targeting the FPGA ZCU208 for fixed-point precision simulation. The MATLAB DSP included resampling from 6.4 to 4sa/sy, coarse frequency and IQ imbalance compensation, resampling to 2 sa/sy for skew correction in the frequency domain, and matched filtering. Timing recovery was applied at 1 sa/sy, followed by the time-domain CMA equalizer with 5 complex taps for each position of the 2×2 MIMO, updated every two symbols. Subsequent steps included frequency offset estimation, phase noise compensation via BPS, a DD-LMS equalizer, and a block for communication performance metrics calculation. The fixed-point CMA equalizer followed the rounding strategy in \cite[Fig.1]{Chin2012}, tested at bit widths $W$ = 5, 6, 7, and 8, with a 2-bit integer part and varying fractional part. The MIMO was implemented via High-Level Synthesis (HLS) in Vitis 2025.1 for FPGA simulation and RTL synthesis. Results were compared to a floating-point MATLAB implementation, serving as a high-resolution benchmark.

\begin{figure*}[t!]
    \centering
    \includegraphics[width=0.99\textwidth]{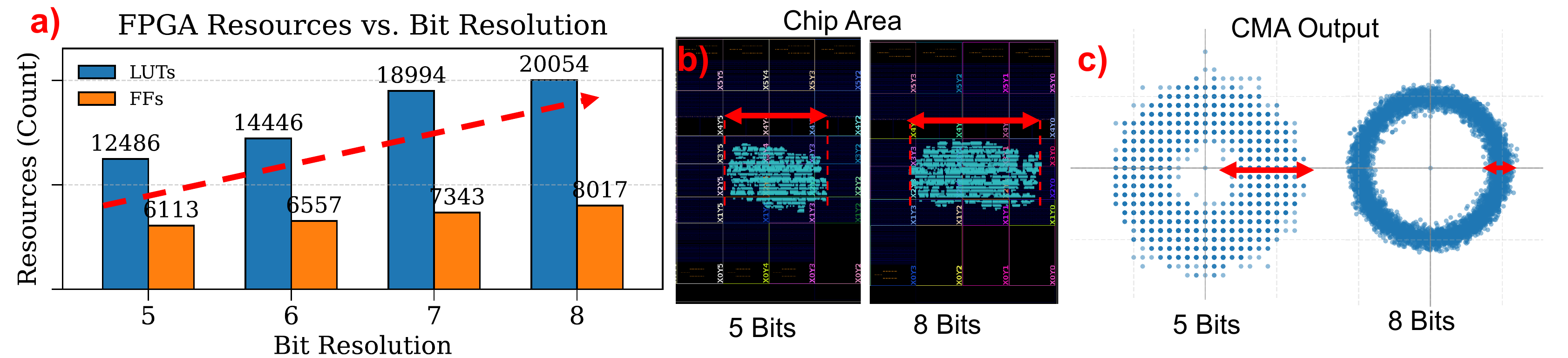}
    \captionsetup{font=footnotesize}
    \caption{a) FPGA resources used for each bit precision b) FPGA chip area required for 5 and 8 bits resolution. c) FPGA CMA output for 5 and 8 bits of precision. With 8 bits precision the CMA output is the best considering convergence to unity modulus}\label{Block_Diagram_Communication_Setup}
    \vspace{-5mm}
\end{figure*}

The coefficients of the MIMO equalizer are a time-varying inverse Jones matrix that conveys the vibration information. By applying a constant input Jones vector to this matrix, we obtain a time-varying output Jones vector. This vector is then converted into a normalized Stokes vector, representing a trajectory on the Poincaré sphere. When the signal amplitude is sufficiently small, the centroid of this trajectory—corresponding to the time-averaged Stokes vector—can be computed. Finally, we rotate this centroid to the north pole of the sphere and apply the same rotation to all Stokes time samples.  By doing this, we can focus the analysis on $S_1$ and $S_2$ and filter out their DC components, which is the same approach used in \cite{Mecozzi2021}. For assessing the communication performance we first assessed the root mean square error vector magnitude (EVM$_{RMS}$) between the recovered signal after the whole described DSP chain and the closest QPSK symbol using \cite[Eq.10]{Shafik2006}, then estimated the BER using  \cite[Eq.13]{Shafik2006} to obtain the estimated BER, which was converted to Q-factor. For the sensing, since the trajectory lies on the Poincaré sphere, we used the metric angular root mean squared error (angular RMSE) calculated by the equation $\text{RMSE}_\theta =  \frac{180}{\pi} \sqrt{\frac{1}{N} \sum_{i=1}^{N} \left[ \cos^{-1}\!\left( \mathbf{S}_{\text{FPP}, i} \cdot \mathbf{S}_{\text{W}, i} \right) \right]^2 }$, in which $S_{FPP}$ and $S_W$ are the stokes vectors for float point precision (FPP) and for different bit precision $W$, respectively. This metric takes into account the angular deviations for each stokes vector from the FPGA compared with the float point precision, used as reference.

\begin{figure*}[b]
    \centering
    \includegraphics[width=1\textwidth]{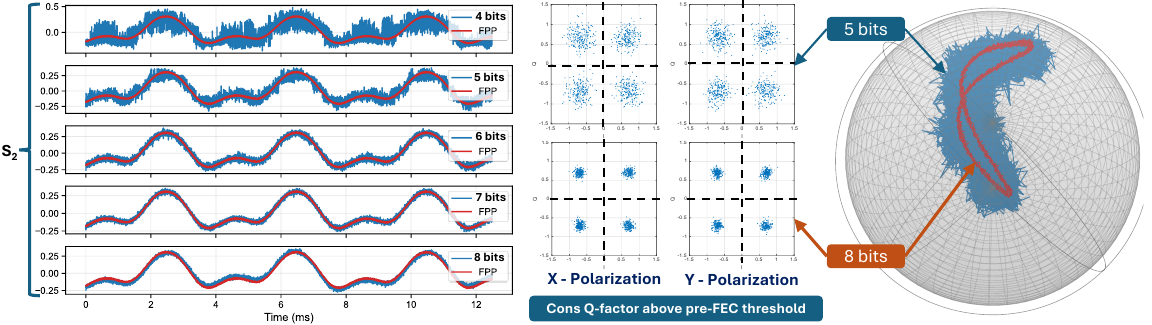}
    \captionsetup{font=footnotesize}
    \caption{Waveform for stokes parameter $S_2$ for each value of bit precision, the quantization-like noise decreases with increasing bit precision. Poincaré sphere trajectory for 5-bit precision (communication optimum \cite{Chin2012}) and for 8-bit precision (minimum quantization-like noise). Increasing bit precision leads to better-defined SOP trajectories.}
    \label{poincare_signal}
\end{figure*}
\vspace{-2mm}

\section{Results and discussion}

\vspace{-1mm}

The number of hardware resources (LUT and FF) scaled linearly according to the number of bits used after synthesis of the CMA equalizer as depicted in Fig. \ref{Block_Diagram_Communication_Setup} a). All the operations in the equalizer were implemented only using FF and LUTs, including the multiplications, for a better comparison of hardware complexity across the different implementations. As the FPGA implementation process 5 samples of both polarizations in parallel, the required chip area also increases with bit-width, as seen in b). In c), the CMA output exhibits larger variations due to coarser quantization, which causes filter coefficients to oscillate around their expected values during symbol-rate updates, resulting in constant transitions between levels during the coefficient updates. Since the SOP is computed from these coefficients, the resulting time-domain SOP signals show increasing noise as the bit-width decreases, as illustrated in Fig. \ref{poincare_signal}. Consequently, the derived Stokes parameters also exhibit higher noise. Therefore, superimposing the SOP trajectories on the Poincaré sphere for $W = 5$ and $W = 8$ further demonstrates that increasing $W$ improves the SOP signal quality and reduces noise.

\begin{figure*}[t]
    \centering
    \includegraphics[width=1\textwidth]{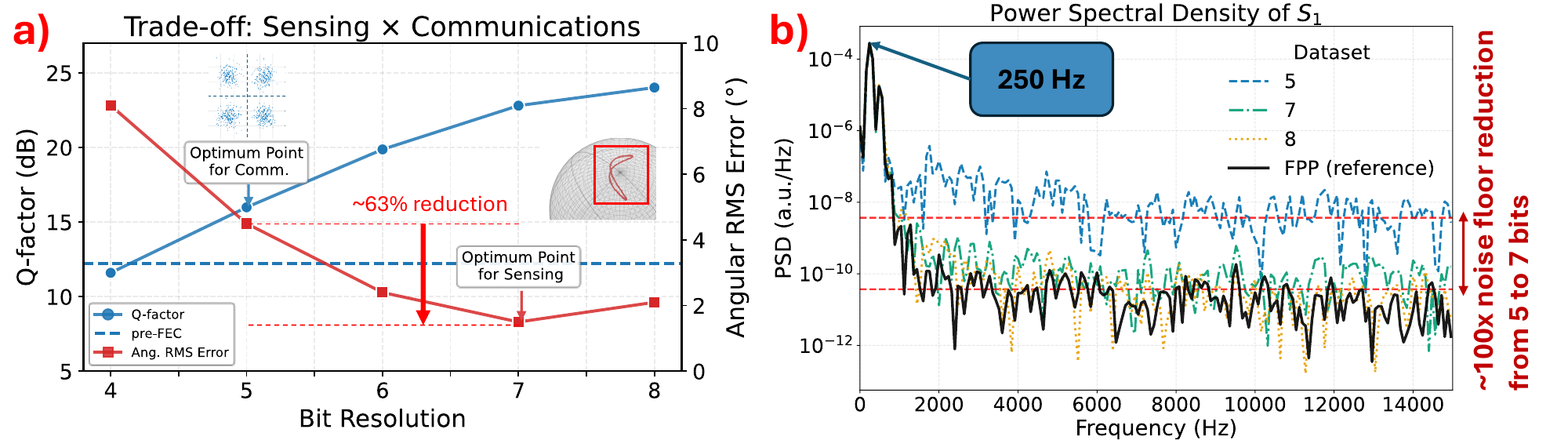}
    \vspace{-7mm}
    \captionsetup{font=footnotesize}
    \caption{Angular RMS error and Q-Factor for each value of bit precision, minimum value of angular RMSE for 7-bit precision. Power spectral density showing the fundamental frequency of the vibration (250 Hz) for $S_1$ and the quantization-like noise in $S_1$ signal for 5 and 7-bit precision, the noise drops sharply as the bit resolution increases.}
    \label{angular_error}
    \vspace{-5mm}
\end{figure*}



For a quantitative analysis, Fig. \ref{angular_error} presents two key metrics: (a) the angular RMSE and (b) the PSD of $S_1$, where the 250 Hz vibration is visible. The angular RMSE plot illustrates the sensing–communication trade-off. For bit widths $W \geq 5$, the communication performance is already above the 12.6dB pre-FEC threshold ($BER = 2.18\cdot10^{-5}$ for 2.7\% overhead\cite{Agrell2018}), indicating that increasing bit precision does not improve communication. However, higher bit-widths significantly enhance sensing accuracy: increasing $W$ from 5 to 7 reduces the angular RMSE by 63\%. The PSD plot further confirms this improvement: the quantization noise floor decreases considerably with higher precision, with $W = 7$ achieving a $\sim$100× reduction compared to $W = 5$. This demonstrates that increasing bit precision substantially lowers the noise floor, enhancing the sensitivity of the SOP-based sensor. These results collectively show that while communication saturates at moderate precision, sensing performance continues to benefit from higher CMA bit-widths.


%
%


\vspace{-2mm}
\section{Conclusion}
\vspace{-1mm}
This work examined the effect of CMA bit precision on state-of-polarization (SOP) sensing. We showed that, while a 5-bit precision is sufficient to achieve a performance above the pre-FEC threshold, the angular RMS error relative to the floating-point SOP decreases 63\% as the precision is increased to 7 bits. In addition, this resolution increment reduced the noise floor $\sim$100×, which enables spectral features to be more clearly resolved and the sensor's sensitivity to be improved. However, higher precision increases FPGA area and resource usage. These findings underscore the importance of the equalizer's fixed-point resolution in SOP signal quality and provide guidance in designing coherent receivers with integrated SOP measurement capabilities.

\vspace{2pt}{\footnotesize
\noindent\textbf{Acknowledgements:} 
This work is supported by EPSRC grant TRANSNET (EP/R035342/1), DSIT and RAEng under Research Fellowship.}

\end{document}